\documentclass[twocolumn,letterpaper]{article}
\usepackage{usenix}
\usepackage{amsmath}
\usepackage{amssymb}
\usepackage{amsfonts}
\usepackage{setspace}
\usepackage{cite}
\usepackage{ifthen}
\usepackage{epsfig}
\usepackage{array}
\usepackage{subfig}
\usepackage{color}
\usepackage{caption}
\usepackage[table]{xcolor}
\usepackage{xspace}
\usepackage{multicol}
\usepackage{mathdots}
\usepackage[textwidth=6.5in,textheight=9in]{geometry}
\usepackage{authblk}
\usepackage{enumitem}
\usepackage{titlesec}
\titlespacing\section{0pt}{8pt plus 1pt minus 1pt}{6pt plus 1pt minus 1pt}
\titlespacing\subsection{0pt}{8pt plus 1pt minus 1pt}{6pt plus 1pt minus 1pt}
\titlespacing\subsubsection{0pt}{8pt plus 1pt minus 1pt}{6pt plus 1pt minus 1pt}

\newtheorem{example}{Example}

\newcommand{\beq}{\begin{equation}}
\newcommand{\eeq}{\end{equation}}
\newcommand{\bea}{\begin{eqnarray}}
\newcommand{\eea}{\end{eqnarray}}

\newcommand{\clustera}{Facebook's warehouse cluster\xspace}
\newcommand{\newcodename}{Piggybacked-RS\xspace}
\newcommand{\numparity}{r}

\newcommand{\clusterone}{Cluster~A\xspace}

\title{Impact of Erasure-codes recovery on network infrastructure: A study and proposed solution}

\title{\vspace{-3cm}Impact of erasure-coding recovery on the network: A study on Facebook warehouse cluster and a proposed solution}
\title{\vspace{-3cm}Network Effects of Erasure-codes in Data Centers: A Study on the Facebook Warehouse Cluster and a Proposed Solution\vspace{-.3cm}}
\title{\vspace{-2.3cm}A Solution to the Network Challenges of Data Recovery in Erasure-coded Distributed Storage Systems: A Study on the Facebook Warehouse Cluster\vspace{-.3cm}}

\author[1]{K. V. Rashmi}
\author[1]{Nihar B. Shah}
\author[2]{Dikang Gu}
\author[2]{Hairong Kuang}
\author[2]{Dhruba Borthakur}
\author[1]{Kannan Ramchandran} 
\affil[1]{UC Berkeley}\affil[2]{Facebook}

\begin{document}
\pagestyle{empty}
\maketitle
\vspace*{-2cm}



\begin{abstract}
Erasure codes, such as Reed-Solomon (RS) codes, are being increasingly employed in data centers to combat the cost of reliably storing large amounts of data. Although these codes provide optimal storage efficiency, they require significantly high network and disk usage during recovery of missing data.

In this paper, we first present a study on the impact of recovery operations of erasure-coded data on the data-center network, based on measurements from Facebook's warehouse cluster in production. To the best of our knowledge, this is the first study of its kind available in the literature. Our study reveals that recovery of RS-coded data results in a significant increase in network traffic, more than a hundred terabytes per day, in a cluster storing multiple petabytes of RS-coded data.

To address this issue, we present a new storage code using our recently proposed \textit{Piggybacking} framework, that reduces the network and disk usage during recovery by $30\%$ in theory, while also being storage optimal and supporting arbitrary design parameters. The implementation of the proposed code in the Hadoop Distributed File System (HDFS) is underway. We use the measurements from the warehouse cluster to show that the proposed code would lead to a reduction of close to fifty terabytes of cross-rack traffic per day. 
\end{abstract}

\section{Introduction}\label{sec:intro}
Data centers today typically employ commodity components for cost considerations. These individual components are unreliable, and as a result, the system must deal with frequent failures of these components. In addition, various additional issues such as software glitches, machine reboots and maintenance operations also contribute to machines being rendered unavailable from time to time. In order to ensure that the data remains reliable and available even in the presence of frequent machine-unavailability, data is replicated across multiple machines, typically across multiple racks as well. For instance, the Google File System~\cite{ghemawat2003google_short} and the Hadoop Distributed File System (HDFS)~\cite{borthakur2008hdfs} store three copies of all data by default. Although disk storage seems cheap for small amounts of data, the massive scales of operation of today's data-centers make storing multiple copies an expensive option. As a result, several large-scale distributed storage systems~\cite{ghemawat2003google_short,hdfs_codes_blog} now employ erasure codes that provide significantly higher storage efficiency, with the most popular choice being the Reed-Solomon (RS) code~\cite{reedSolomon_supershort}.

\begin{figure}[t]
\centering
\vspace{-1cm}
\includegraphics[width=0.45\textwidth]{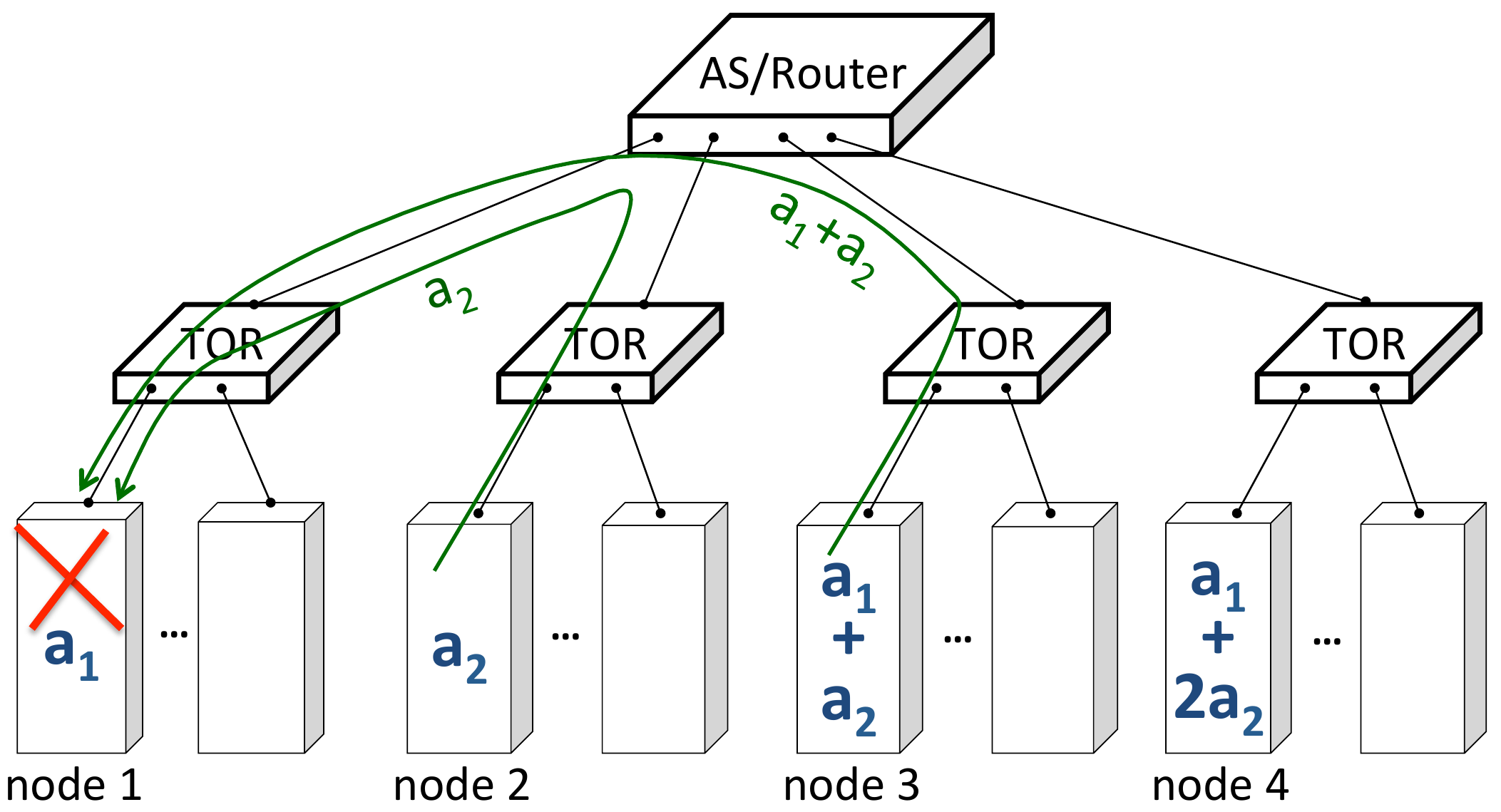}
\caption{Increase in network usage during recovery of erasure-coded data: recovering a single missing unit ($a_1$) requires transfer of two units through the top-of-rack (TOR) switches and the aggregation switch (AS).}
\label{fig:intro}
\end{figure}

An RS code is associated with two parameters: $k$ and $\numparity$. A $(k,\ r)$ RS code encodes $k$ units of data into $\numparity$ \textit{parity} units, in a manner that all the $k$ data units are recoverable from any $k$ out of these $(k+\numparity)$ units. It thus allows for tolerating failure of any $\numparity$ of its $(k+\numparity)$ units. The collection of these $(k+\numparity)$ units together is called a \textit{stripe}. In a system employing an RS code, the data and the parity units belonging to a stripe are stored on different machines to tolerate maximum unavailability. In addition, these machines are chosen to be on different racks to tolerate rack failures as well. An example of such a setting is depicted in Fig.~\ref{fig:intro}, with an RS code having parameters $(k=2,r=2)$. Here $\{a_1,\ a_2\}$ are the two data units, which are encoded to generate two parity units, $(a_1 + a_2)$ and $(a_1 + 2a_2)$. The figure depicts these four units stored across four nodes (machines) in different racks.

Two primary reasons that make RS codes particularly attractive for large-scale distributed storage systems are: (a) they are storage-capacity optimal, i.e., a $(k,\ \numparity)$ RS code entails the minimum storage overhead among all $(k, \ \numparity)$ erasure codes that tolerate any $\numparity$ failures,\footnote{In the parlance of coding theory, an RS code has the property of being `Maximum Distance Separable (MDS)'.} (b) they can be constructed for arbitrary values of the parameters $(k, \ \numparity)$, thus allowing complete flexibility in the choice of these parameters. For instance, the warehouse cluster at Facebook employs an RS code with parameters $(k=10,\ r=4)$, thus resulting in a $1.4\times$ storage requirement, as compared to $3\times$ under conventional replication, for a similar level of reliability.

While deploying RS codes in data centers improves storage efficiency, it however results in a significant increase in the disk and  network bandwidth usage. This phenomenon occurs due to the considerably high download requirement during recovery of any missing unit, as elaborated below. In a system that performs replication, a missing data unit can be restored simply by copying it from another existing replica. However, in an RS coded system, no such replica exists. To see the recovery operation under an RS code, let us first consider the example $(k=2,\ r=2)$ RS code in Fig.~\ref{fig:intro}. The figure illustrates the recovery of the first data unit $a_1$ (node $1$) from nodes $2$ and $3$. Observe that this recovery operation requires the transfer of two units across the network. In general, under a $(k,\ r)$ RS code, recovery of a single unit involves the download of some $k$ of the remaining units. An amount equal to the logical size of the data in the stripe is thus read and downloaded, from which the required missing unit is recovered. 


The contributions of this paper are divided into two parts. The first part of the paper presents measurements from Facebook's warehouse cluster in production that stores hundreds of petabytes of data across a few thousand machines, studying the impact of recovery operations of RS-coded data on the network infrastructure. The study reveals that there is a significant increase in the cross-rack traffic due to the recovery operations of RS-coded data, thus significantly increasing the burden on the already oversubscribed TOR switches. To the best of our knowledge, this is the first study available in the literature which looks at the effect of recovery operations of  erasure codes on the network usage in data centers. The second part of the paper describes the design of a new code that reduces the disk and network bandwidth consumption by approximately $30\%$ (in theory), and also retains the two appealing properties of RS codes, namely storage optimality and flexibility in the choice of code parameters. 
As a proof-of-concept, we use measurements from the cluster in production to show that employing the proposed new code can indeed lead to a significant reduction in cross-rack network traffic. 

The rest of the paper is organized as follows. Section~\ref{sec:measurements} presents measurements from the Facebook warehouse cluster, and an analysis of the impact of recovery of RS-coded data on the network infrastructure. 
Section~\ref{sec:code} presents the design of the new `Piggybacked-RS' code, along with a discussion on its expected performance. Section~\ref{sec:current} describes the current state and future plans for this project. Section~\ref{sec:literature} discusses related works. 

\section{Measurements from \clustera}~\label{sec:measurements}
\vspace{-1cm}
\subsection{Brief description of the system}
\begin{figure}[t]
\centering
\includegraphics[width=0.48\textwidth]{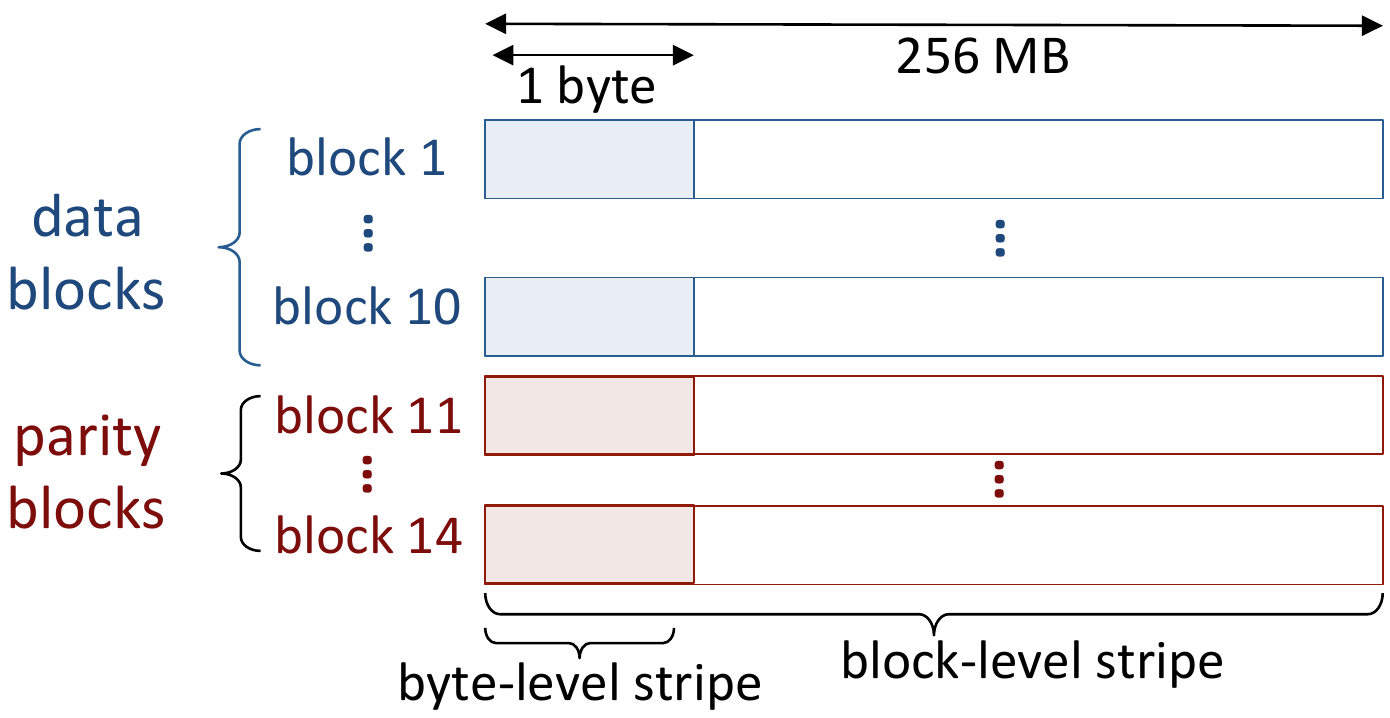}
\caption{Erasure coding across blocks: $10$ data blocks encoded using $(10, \ 4)$ RS code to generate $4$ parity blocks.}
\label{fig:blocks}
\end{figure}

The warehouse cluster comprises of two HDFS clusters, which we shall refer to as clusters A and B. In terms of physical size, these clusters together store hundreds of petabytes of data, and the storage capacity used in each cluster is growing at a rate of a few petabytes every week. These clusters store data across a few thousand machines, each of which has a storage capacity of $24$-$36TB$. The data stored in these clusters is immutable until it is deleted, and is compressed prior to being stored in the cluster. Map-reduce jobs are the predominant consumers of the data stored in the cluster.

\begin{figure*}[t!]
\centering
\subfloat[]{
\!\!\includegraphics[trim=.1in .03in .5in .3in,clip,width=.4\textwidth]{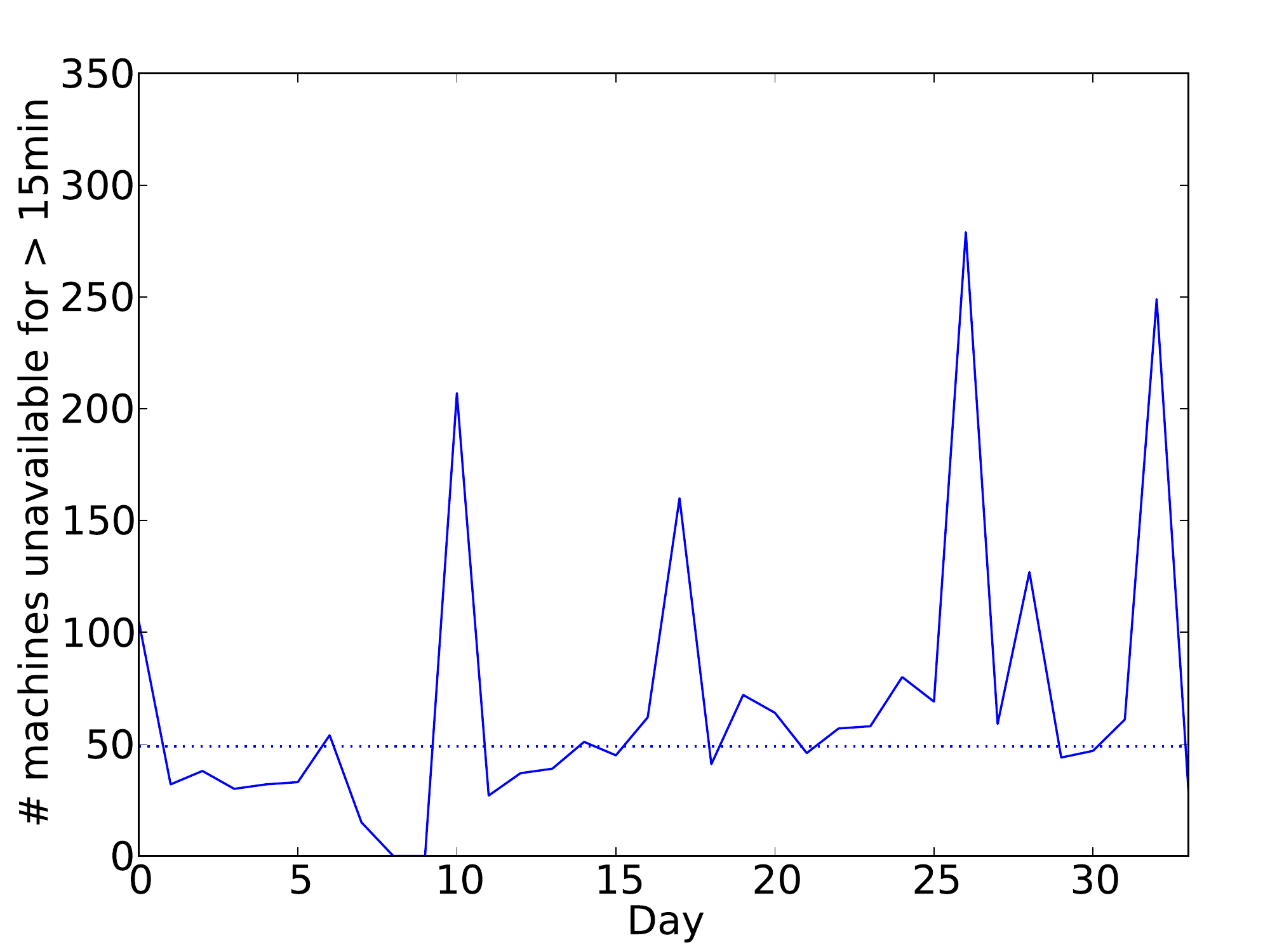}
\label{fig:failcounts}
}~~~~~~~~
\subfloat[]{
\includegraphics[trim=.65in .75in .2in 1.01in,clip,width=.52\textwidth]{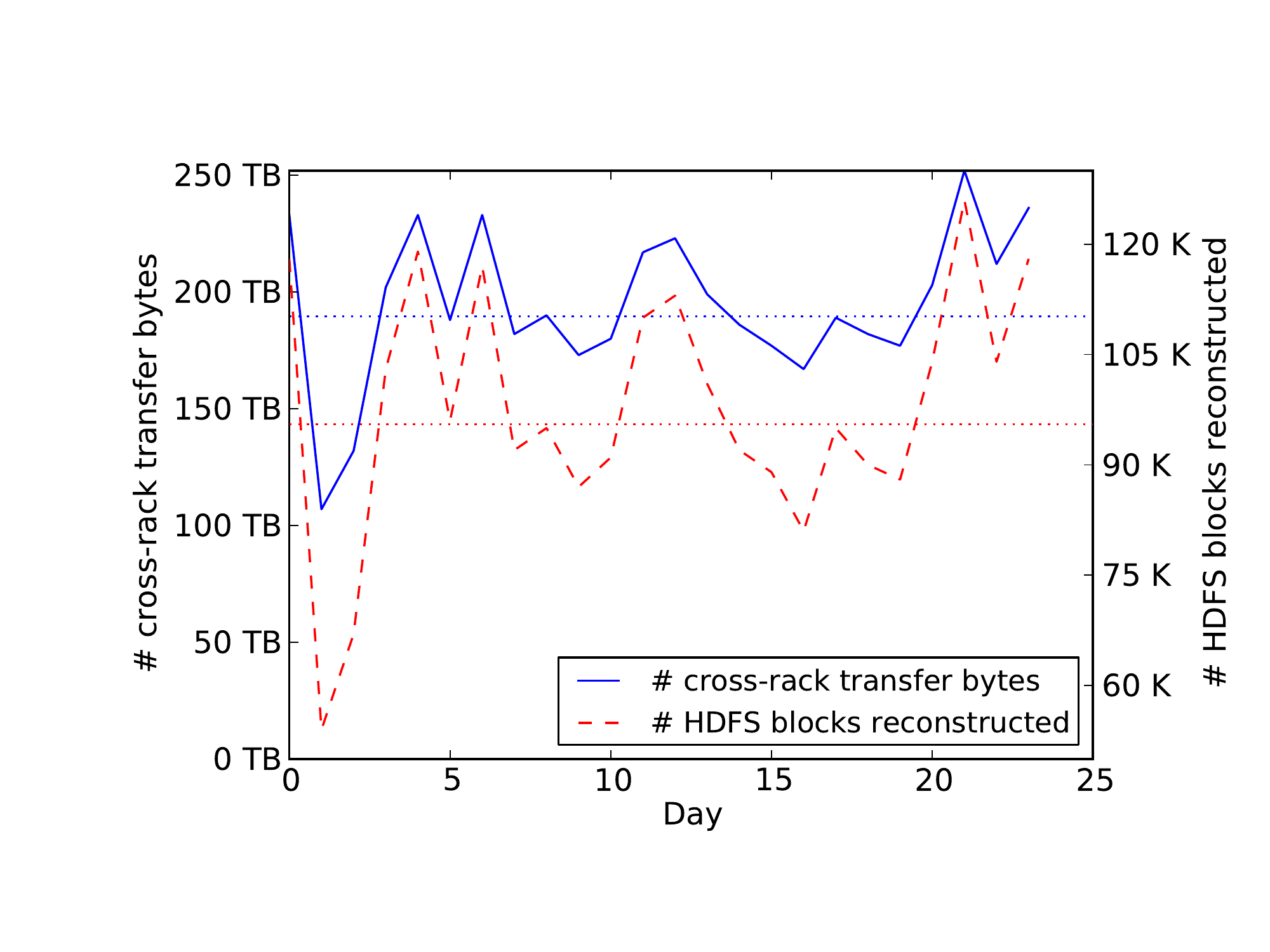}
\label{fig:crossrackbytes}
}
\vspace{-.2cm}
\caption{Measurements from Facebook's warehouse cluster: \protect\subref{fig:failcounts} the number of machines unavailable for more than 15 minutes in a day, over a duration of $3$ months, and \protect\subref{fig:crossrackbytes} RS-coded HDFS blocks reconstructed and cross-rack bytes transferred for recovery operations per day, over a duration of around a month. The dotted lines in each plot represent the median values.}%
\end{figure*}
Since the amount of data stored is very large, the cost of operating the cluster is dominated by the cost of the storage capacity. The most frequently accessed data is stored as $3$ replicas, to allow for efficient scheduling of the map-reduce jobs. In order to save on the storage costs, the data which has not been accessed for more than three months is stored as a $(10,4)$ RS code. The two clusters together store more than ten petabytes of RS-coded data. Since the employed RS code has a redundancy of only $1.4$, this results in huge savings in storage capacity as compared to $3$-way replication.


We shall now delve deeper into details of the RS-coded system and the recovery process. A file or a directory is first partitioned into blocks of size $256$MB. These blocks are grouped into sets of $10$ blocks each; every set is then encoded with a $(10, \ 4)$ RS code to obtain $4$ parity blocks. As illustrated in Fig.~\ref{fig:blocks}, one byte each at corresponding locations in the $10$ data blocks are encoded to generate the corresponding bytes in the $4$ parity blocks. The set of these $14$ blocks constitutes a \textit{stripe} of blocks. The $14$ blocks belonging to a particular stripe are placed on $14$ different (randomly chosen) machines. In order to secure the data against rack-failures, these machines are chosen from different racks. To recover a missing block, any $10$ of the remaining $13$ blocks of its stripe are downloaded. Since each block is placed on a different rack, these transfers take place through the TOR switches. This consumes precious cross-rack bandwidth that is heavily oversubscribed in most data centers including the one studied here. 

As discussed above, the data to be encoded is chosen based on its access pattern. We have observed that there exists a large portion of data in the cluster which is not RS-encoded at present, but has access patterns that permit erasure coding. The increase in the load on the already oversubscribed network infrastructure, resulting from the recovery operations, is the primary deterrent to the erasure coding of this data. 


\subsection{Data-recovery in erasure-coded systems: Impact on the network}\label{sec:plots}
We have performed measurements on \clustera to study the impact of the recovery operations of the erasure-coded data. An analysis of these measurements reveals that the large downloads for recovery required under the existing codes is indeed an issue. We present some of our findings below. 

\begin{enumerate}
\item \textit{Unavailability Statistics:} We begin with some statistics on machine unavailability. Fig.~\ref{fig:failcounts} plots the number of machines unavailable for more than $15$ minutes in a day, over the period $22^{\text{nd}}$ Jan. to $24^{\text{th}}$ Feb. $2013$ ($15$ minutes is the default wait-time of the cluster to flag a machine as unavailable). The median is more than $50$ machine-unavailability events per day. This reasserts the necessity of redundancy in the data for both reliability and availability. A subset of these events ultimately trigger recovery operations. 

\item \textit{Number of missing blocks in a stripe:} Of all the stripes that have one or more blocks missing, on an average, $98.08\%$ have exactly one block missing. The percentage of stripes with two blocks missing is $1.87\%$, and with three or more blocks missing is $0.05 \%$. Thus recovering from single failures is by-far the most common scenario. This is based on data collected over a  period of $6$ months.

\end{enumerate}

\noindent We now move on to measurements pertaining to recovery operations for RS-coded data. The analysis is based on the data collected from \clusterone for the first $24$ days of Feb. $2013$.

\begin{enumerate}
\setcounter{enumi}{2}
\item \textit{Number of block-recoveries:} 
Fig.~\ref{fig:crossrackbytes} shows the number of block recoveries triggered each day. A median of $95,500$ blocks of RS-coded data are required to be recovered each day. 



\item \textit{Cross-rack bandwidth consumed:} 
We measured the number of bytes transferred across racks for the recovery of RS-coded blocks. The measurements, aggregated per day, are depicted in Fig.~\ref{fig:crossrackbytes}. As shown in the figure, a median of more than $180TB$ of data is transferred through the TOR switches every day for RS-coded data recovery. Thus the recovery operations consume a large amount of cross-rack bandwidth, thereby rendering the bandwidth unavailable for the foreground map-reduce jobs.
\vspace{-.2cm}
\end{enumerate}

This study shows that employing traditional erasure-codes such as RS codes puts a massive strain on the network infrastructure due to their inefficient recovery operations. This is the primary impediment towards a wider deployment of erasure codes in the clusters. We address this concern in the next section by designing codes that support recovery with a smaller download. 




\section{Piggybacked-RS codes $\&$ system design}~\label{sec:code}
In this section, we present the design of a new family of codes that address the issues related to RS codes discussed in the previous section, while retaining the storage optimality and flexibility in the choice of parameters. These codes, which we term the \textit{\newcodename} codes, are based on our recently proposed \textit{Piggybacking} framework~\cite{piggyISIT2013_short}. 

\subsection{Code design}
\begin{figure}[t]
\centering
\includegraphics[width=.33\textwidth]{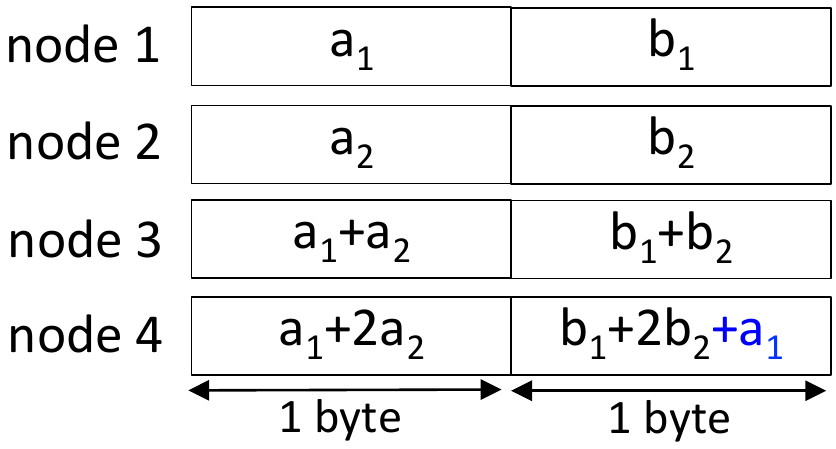}
\caption{An illustration of the idea of \textit{piggybacking}: a toy example of piggybacking a $(k=2,\ r=2)$ RS code.}
\label{fig:piggy_example}
\end{figure}

A \newcodename code is constructed by taking an existing RS code and adding carefully designed functions of one byte-level stripe onto the parities of other byte-level stripes (recall Fig.~\ref{fig:blocks} for the definition of \textit{byte-level stripes}). The functions are designed in a manner that reduces the amount of read and download required during recovery of individual units, while retaining the storage optimality. We illustrate this idea through a toy example which is depicted in Fig.~\ref{fig:piggy_example}.

\begin{example}\label{eg:piggy}
Let $k=2$ and $r=2$. Consider two sets of data units $\{a_1,a_2\}$ and $\{b_1,b_2\}$ corresponding to two byte-level stripes. We first encode the two data units in each of the two stripes using a $(k=2,\ r=2)$ RS code, and then add $a_1$ from the first stripe to the second parity of the second stripe as shown in Fig.~\ref{fig:piggy_example}. Now, recovery of node $1$ can be carried out by downloading $b_2$, $(b_1+b_2)$ and $(b_1+2b_2+a_1)$ from nodes $2$, $3$ and $4$ respectively. The amount of download is thus $3$ bytes, instead of $4$ as previously under the RS code. Thus adding the piggyback, $a_1$, aided in reducing the amount of data read and downloaded for the recovery of node $1$. One can easily verify that this code can tolerate the failure of any $2$ of the $4$ nodes, and hence retains the fault tolerance and storage efficiency of the RS code. Finally, we note that the download during recovery is reduced without requiring any additional storage. 
\end{example}

We propose a $(10, \ 4)$ piggybacked-RS code as an alternative to the $(10, \ 4)$ RS code employed in HDFS. The construction of this code generalizes the idea presented in Example~\ref{eg:piggy}: two byte-level stripes are encoded together and specific functions of the first stripe are added to the parities of the second stripe. This code, in theory, saves around $30\%$ on average in the amount of read and download for recovery of single block  failures. Furthermore, like the RS code, this code is storage optimal and can tolerate any $4$ failures in a stripe. Due to lack of space, details on the code construction are omitted, and the reader is referred to~\cite{piggyISIT2013_short} for a description of the general Piggybacking framework.

\subsection{Estimated performance}~\label{sec:pof}
We are currently in the process of implementing the proposed code in HDFS. The expected performance of this code in terms of the key metrics of amount of download, time for recovery, and reliability is discussed below.

\textit{Amount of download:} As discussed in Section~\ref{sec:plots}, $98\%$ of the block recovery operations correspond to the case of single block recovery in a stripe. For this case, the proposed \newcodename code reduces the disk and network bandwidth requirement by $30\%$. Thus from the measurements presented in Section~\ref{sec:plots}, we estimate that replacing the RS code with the \newcodename code would result in a reduction of more than $50TB$ of cross-rack traffic per day. This is a significant reduction which would allow for storing a greater fraction of data using erasure codes, thereby saving storage capacity.


\textit{Time taken for recovery:} 
Recovering a missing block in a system employing a $(k, \ r)$ RS code requires connecting to only $k$ other nodes. On the other hand, efficient recovery under Piggybacked-RS codes necessitate connecting to more nodes, but requires the download of a smaller amount of data in total. We have conducted preliminary experiments in the cluster which indicate that connecting to more nodes does not affect the recovery time in the cluster. At the scale of multiple megabytes, the system is limited by the network and disk bandwidths, making the recovery time dependent only on the total amount of data read and transferred. The \newcodename code reduces the total amount of data read and downloaded, and thus is expected to lower the recovery times. 

\textit{Storage efficiency and reliability:} The Piggybacked-RS code retains the storage efficiency and failure handling capability of the RS code: it  does not require any additional storage and can tolerate any $r$ failures in a stripe. Moreover, as discussed above, we believe that the time taken for recovery of a failed block will be lesser than that in RS codes. Consequently, we believe that the mean time to data loss (MTTDL) of the resulting system will be higher than that under RS codes.

\section{Current state of the project}\label{sec:current}
We are currently implementing \newcodename codes in HDFS, and upon completion, we plan to evaluate its performance on a production-scale cluster. Moreover, we are continuing to collect measurements, including metrics in addition to those presented in this paper.

\section{Related work}\label{sec:literature}
There have been several studies on failure statistics in storage systems, e.g., see~\cite{disk_real_MTTF_short,ford2010availability_short} and the references therein.  In~\cite{weatherspoon2002erasure_short}, the authors perform a theoretical comparison of replication and erasure-codes for peer-to-peer storage systems. However, the setting considered therein does not take into account the fact that recovering a single block in a $(k, \ r)$ RS code requires $k$ times more data to be read and downloaded. On the other hand, the primary focus of the present paper is on analysing the impact of this excess bandwidth consumption. To the best of our knowledge, this is the first work to study the impact of recovery operations of erasure codes on the network usage in data centers.

The problem of designing codes to reduce the amount of download for recovery has received quite some theoretical interest in the recent past. The idea of connecting to more nodes and downloading smaller amounts of data from each node was proposed in~\cite{YunDimKanJournal_supershort} as a part of the `regenerating codes model', along with the lower bounds on the amount of download. However, existing constructions of regenerating codes either require a high redundancy~\cite{ourProductMatrix_supershort} or support at most $3$ parities~\cite{tamo2011mds_supershort, papailiopoulos2011repair_short,cadambe2011subspace}. 
Rotated-RS~\cite{khan2012rethinking_short} is another class of codes proposed for the same purpose. However, it supports at most $3$ parities, and furthermore its fault tolerance is established via a computer search. Recently, optimized recovery algorithms~\cite{arrayrepair2_dimakis_vshort,xiang2010optimal_short} have been proposed for EVENODD and RDP codes, but they support only $2$ parities. For the parameters where~\cite{khan2012rethinking_short,arrayrepair2_dimakis_short,xiang2010optimal_short} exist, we have shown in~\cite{piggyISIT2013_short} that the Piggybacked-RS codes perform at least as well. Moreover, the \newcodename codes support an arbitrary number of parities. 

In~\cite{huang2012erasure_short,asterisxoring}, a new class of codes called LRCs are proposed that reduce the disk and network bandwidth requirement during recovery. In~\cite{huang2012erasure_short}, the authors also provide measurements from Windows Azure Storage showing the reduction in read latency for missing blocks when LRCs are employed; however, no system measurements regarding bandwidth are provided. In~\cite{asterisxoring}, the authors perform simulations with LRCs on Amazon EC2, where they show reduction in latency and recovery bandwidth. Although LRCs reduce the bandwidth consumed during recovery, they are not storage efficient: LRCs reduce the amount of download by storing additional parity blocks, whereas \newcodename do not require any additional storage. In the parlance of coding theory, \newcodename codes (like RS codes) are MDS and are hence storage optimal, whereas LRCs are not.

\small
\begin{spacing}{0.9}
\bibliographystyle{IEEEtran}
\vspace{-.1cm}
\bibliography{../bibtex/distributedStorage}

\begin{thebibliography}{10}
\providecommand{\url}[1]{#1}
\csname url@samestyle\endcsname
\providecommand{\newblock}{\relax}
\providecommand{\bibinfo}[2]{#2}
\providecommand{\BIBentrySTDinterwordspacing}{\spaceskip=0pt\relax}
\providecommand{\BIBentryALTinterwordstretchfactor}{4}
\providecommand{\BIBentryALTinterwordspacing}{\spaceskip=\fontdimen2\font plus
\BIBentryALTinterwordstretchfactor\fontdimen3\font minus
  \fontdimen4\font\relax}
\providecommand{\BIBforeignlanguage}[2]{{%
\expandafter\ifx\csname l@#1\endcsname\relax
\typeout{** WARNING: IEEEtran.bst: No hyphenation pattern has been}%
\typeout{** loaded for the language `#1'. Using the pattern for}%
\typeout{** the default language instead.}%
\else
\language=\csname l@#1\endcsname
\fi
#2}}
\providecommand{\BIBdecl}{\relax}
\BIBdecl

\bibitem{ghemawat2003google_short}
S.~Ghemawat, H.~Gobioff, and S.~Leung, ``The google file system,'' in \emph{ACM
  SOSP}, 2003.

\bibitem{borthakur2008hdfs}
\BIBentryALTinterwordspacing
D.~Borthakur, ``{HDFS architecture guide},'' 2008. [Online]. Available:
  \url{http://hadoop. apache. org/common/docs/current/hdfs design. pdf}
\BIBentrySTDinterwordspacing

\bibitem{hdfs_codes_blog}
\BIBentryALTinterwordspacing
------, ``{HDFS and Erasure Codes (HDFS-RAID)},'' 2009. [Online]. Available:
  \url{http://hadoopblog.blogspot.com/2009/08/hdfs-and-erasure-codes-hdfs-raid.html}
\BIBentrySTDinterwordspacing

\bibitem{reedSolomon_supershort}
I.~Reed and G.~Solomon, ``Polynomial codes over certain finite fields,''
  \emph{Journal of SIAM}, 1960.

\bibitem{piggyISIT2013_short}
K.~V. Rashmi, N.~B. Shah, and K.~Ramchandran, ``A piggybacking design framework
  for read-and download-efficient distributed storage codes,'' in \emph{Proc.
  IEEE ISIT (to appear)}, Jul. 2013.

\bibitem{disk_real_MTTF_short}
B.~Schroeder and G.~Gibson, ``Disk failures in the real world: What does an
  {MTTF} of 1,000,000 hours mean to you?'' in \emph{Proc. FAST}, 2007.

\bibitem{ford2010availability_short}
D.~Ford, F.~Labelle, F.~Popovici, M.~Stokely, V.~Truong, L.~Barroso, C.~Grimes,
  and S.~Quinlan, ``Availability in globally distributed storage systems,'' in
  \emph{OSDI}, 2010.

\bibitem{weatherspoon2002erasure_short}
H.~Weatherspoon and J.~D. Kubiatowicz, ``Erasure coding vs. replication: A
  quantitative comparison,'' in \emph{IPTPS}, 2002.

\bibitem{YunDimKanJournal_supershort}
A.~G. Dimakis, P.~B. Godfrey, Y.~Wu, M.~Wainwright, and K.~Ramchandran,
  ``Network coding for distributed storage systems,'' \emph{IEEE Trans. Inf.
  Th.}, Sep. 2010.

\bibitem{ourProductMatrix_supershort}
K.~V. Rashmi, N.~B. Shah, and P.~V. Kumar, ``Optimal exact-regenerating codes
  for the {MSR} and {MBR} points via a product-matrix construction,''
  \emph{IEEE Trans. Inf. Th.}, Aug. 2011.

\bibitem{tamo2011mds_supershort}
I.~Tamo, Z.~Wang, and J.~Bruck, ``{MDS} array codes with optimal rebuilding,''
  in \emph{IEEE ISIT}, Jul. 2011.

\bibitem{papailiopoulos2011repair_short}
D.~Papailiopoulos, A.~Dimakis, and V.~Cadambe, ``Repair optimal erasure codes
  through {H}adamard designs,'' in \emph{Proc. Allerton Conf.}, 2011.

\bibitem{cadambe2011subspace}
\BIBentryALTinterwordspacing
V.~R. Cadambe, C.~Huang, S.~A. Jafar, and J.~Li, ``Optimal repair of mds codes
  in distributed storage via subspace interference alignment,'' 2011. [Online].
  Available: \url{arXiv:1106.1250}
\BIBentrySTDinterwordspacing

\bibitem{khan2012rethinking_short}
O.~Khan, R.~Burns, J.~Plank, W.~Pierce, and C.~Huang, ``Rethinking erasure
  codes for cloud file systems: minimizing {I/O} for recovery and degraded
  reads,'' in \emph{USENIX FAST}, 2012.

\bibitem{arrayrepair2_dimakis_vshort}
Z.~Wang, A.~G. Dimakis, and J.~Bruck, ``Rebuilding for array codes in
  distributed storage systems,'' in \emph{ACTEMT}, 2010.

\bibitem{xiang2010optimal_short}
L.~Xiang, Y.~Xu, J.~Lui, and Q.~Chang, ``Optimal recovery of single disk
  failure in {RDP} code storage systems,'' in \emph{ACM SIGMETRICS}, 2010.

\bibitem{arrayrepair2_dimakis_short}
Z.~Wang, A.~G. Dimakis, and J.~Bruck, ``Rebuilding for array codes in
  distributed storage systems,'' in \emph{ACTEMT}, Dec. 2010.

\bibitem{huang2012erasure_short}
C.~Huang, H.~Simitci, Y.~Xu, A.~Ogus, B.~Calder, P.~Gopalan, J.~Li, and
  S.~Yekhanin, ``Erasure coding in {Windows Azure} storage,'' in \emph{USENIX
  ATC}, 2012.

\bibitem{asterisxoring}
S.~Mahesh, M.~Asteris, D.~Papailiopoulos, A.~G. Dimakis, R.~Vadali, S.~Chen,
  and D.~Borthakur, ``Xoring elephants: Novel erasure codes for big data,'' in
  \emph{VLDB Endowment (to appear)}, 2013.

\end{thebibliography}
\end{spacing}
\end{document}